\documentclass[%
 aip,
 amsmath,amssymb,
 reprint,%
]{revtex4-1}

\usepackage{graphicx}
\usepackage{dcolumn}
\usepackage{bm}

\usepackage[utf8]{inputenc}
\usepackage[T1]{fontenc}
\usepackage{mathptmx}
\usepackage{etoolbox}
\usepackage{xcolor}

\usepackage{upgreek}



\begin{document}

\title{Rydberg Atom Electric Field Sensors as Linear Time-invariant Systems}
\author{N. Malvania}\thanks{Corresponding Author, email: nmalvania@mitre.org}
\affiliation{The MITRE Corporation, 7515 Colshire Dr.,  McLean, VA 22012, USA}%
\author{G. Jacyna}
\affiliation{The MITRE Corporation, 7515 Colshire Dr.,  McLean, VA 22012, USA}%
\author{B. L. Schmittberger Marlow}
\affiliation{The MITRE Corporation, 7515 Colshire Dr.,  McLean, VA 22012, USA}%
\author{Z. N. Hardesty-Shaw}
\affiliation{The MITRE Corporation, 7515 Colshire Dr.,  McLean, VA 22012, USA}%
\author{K. L. Nicolich}
\affiliation{The MITRE Corporation, 7515 Colshire Dr.,  McLean, VA 22012, USA}%
\author{K. M. Backes}
\affiliation{The MITRE Corporation, 7515 Colshire Dr.,  McLean, VA 22012, USA}%
\author{J. L. MacLennan}
\affiliation{The MITRE Corporation, 7515 Colshire Dr.,  McLean, VA 22012, USA}%
\author{C. T. Fancher}
\affiliation{The MITRE Corporation, 7515 Colshire Dr.,  McLean, VA 22012, USA}%

\begin{abstract}
Over the past decade, Rydberg atom electric field sensors have been under investigation as potential alternatives or complements to conventional antenna-based receivers for select applications in RF communications, remote sensing, and precision metrology. To understand the potential utility of these devices for various use cases, it is crucial to develop models that accurately predict key performance metrics such as instantaneous bandwidth and dynamic range. However, existing numerical models require solving a large set of coupled differential equations that is computationally intensive and lengthy to solve. We present an analytic approach that can be used to derive an impulse response function that allows up to two orders-of-magnitude reduction in computation time compared to the full time-dependent integration of the equations of motion. This approach can be used to enable rapid assessments of the Rydberg sensor's response to various waveforms.

\end{abstract}

\maketitle

The behavior of many physical systems can be understood via their characteristic equations of motion. For complex systems, solving these equations numerically can require lengthy computations. In the case of Rydberg atom electric field sensors, the equations of motion can be used to understand the sensor's response to various waveforms~\cite{PhysRevA.104.032824,PhysRevApplied.18.034030,Schmidt:24} and enhance our ability to understand the sensor's utility for various use cases in RF communications and sensing~\cite{9054945,9374680,SIMONS2021100273}. However, solving these equations numerically for a given RF waveform sometimes requires hours of computational time ~\cite{Meyer:24}, especially when microsecond-level temporal features of the sensor's response must be resolved in millisecond-scale waveforms.

Here we present an analytic perturbative approach to modeling the response of Rydberg atom electric field sensors to incident RF fields that can enable substantial computational speedups. We use this approach to calculate the sensor's impulse response function, which defines how the sensor's output changes in response to an incident RF field. This function can be used and extended to calculate key sensor performance metrics such as instantaneous bandwidth, dynamic range, and sensitivity. In general, the impulse response function of linear time-invariant (LTI) systems can be used to analyze, design, and characterize dynamic situations where the output observables do not depend on the timing of the input. Linear systems can be crucial tools in many physical situations such as circuit analysis and design, feedback and stability in control theory, signal and image processing, and mechanical and fluidic systems among other engineering applications ~\cite{hallauerLTI_2016, delchamps_1988, hespanha_2009}.
~Here we treat the Rydberg atom electric field sensor as an LTI system, which is valid in the regime where the amplitude of the incident RF field is weak enough such that a perturbative analysis is valid. We show that this approach predicts frequency responses that are consistent with those predicted using numerical approaches but can save up to two orders-of-magnitude of computational time.

\begin{figure}[t]
	{
		\centering
		\includegraphics[width=1.0\linewidth]{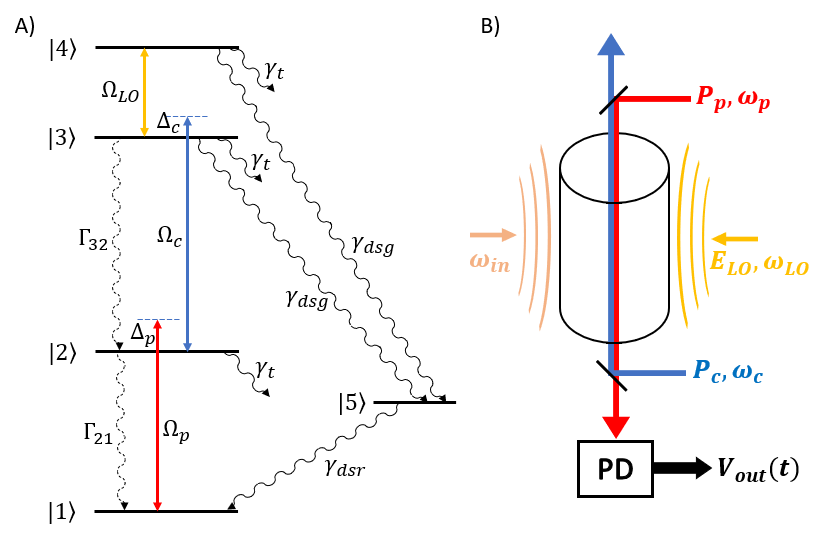}
		\caption[Rydberg Receiver Schematic]{A) 4 level Rydberg sensor model with an additional non-physical ``dummy'' state along with associated Rabi frequencies, detunings, and decay rates. B) Depiction of the measurement setup with probe laser (red), coupling laser (blue), atomic vapor cell, photodetector, RF local oscillator source (yellow), and RF signal field (orange).
		\label{f:RydbergSetup}}
	}
\end{figure}

We consider a Rydberg atom electric field sensor operating with the energy level diagram shown in Fig.~\ref{f:RydbergSetup}A and the heterodyne detection scheme depicted in Fig.~\ref{f:RydbergSetup}B. This heterodyne configuration of a Rydberg atom sensor consists of applying three electromagnetic fields to the atoms\textemdash two counter-propagating optical fields and one RF field\textemdash which together are used to detect a separate weak RF signal field. The two optical fields are referred to as a ``probe field,'' which drives transitions between energy levels $\left|1\right>$ and $\left|2\right>$, and a ``control field,'' which drives transitions between energy levels $\left|2\right>$ and $\left|3\right>$. Together, the two optical fields excite alkali atoms into high-energy Rydberg states via a two-photon process. The locally applied RF field, known as a local oscillator (LO), drives transitions between states $\left|3\right>$ and $\left|4\right>$ and can be used both as a phase reference for detecting phase-modulated fields~\cite{10.1063/1.5088821} and to provide sensitivity enhancements for the sensor\cite{Jing2020,Meyer_2020}. Atoms in Rydberg states possess large electric dipole moments, and an RF signal field passing through the atoms induces perturbations in the atomic energy levels. Any modulations on that RF signal field are transduced onto the optical fields passing through the atomic vapor, and the voltage produced by detecting the transmission of the probe field on a photodetector provides a readout of the baseband signal on the RF field.

Rydberg atom electric field sensors achieve the best sensitivities when the incident RF signal field is close to the frequency spacing between two neighboring Rydberg energy levels. Resonantly coupled Rydberg-Rydberg transitions that are accessible using the two-photon excitation scheme shown in Fig.~\ref{f:RydbergSetup}A typically correspond to RF frequencies in the GHz regime\cite{9374680}. By changing the control laser wavelength, one can access various Rydberg states with different resonantly enhanced RF frequency sensitivities. We note that by using alternative optical excitation schemes~\cite{PhysRevA.107.052605} or local oscillator configurations~\cite{PhysRevA.104.032824}, one can access other frequency bands and achieve near-continuous RF spectral coverage.

In the following subsections we apply linear systems theory to derive the atomic dynamics of a heterodyne-readout based Rydberg atom electric field sensor and present the receiver response and its LTI properties under the influence of a weak RF field. We compare the sensor's frequency response function, which generally takes seconds or minutes to compute, to that obtained by a numerical approach, which can take hours to compute, and we show that results between the two approaches agree. We summarize the primary tradeoffs between the standard numerical technique and our LTI approach in Table~\ref{table1}.

\begin{center}
\begin{table*}[t]
    \begin{tabular}{|p{7.5cm}|p{7.5cm}|}
    \hline
      \textbf{Standard Numerical Solution}   & 
     \textbf{LTI Approach} \\ \hline
       $\checkmark$ Models arbitrary conditions and does not require approximations  & $\times$ ~Method described here can only model the linear regime of the receiver (i.e., RF signals that are weak compared to the local oscillator) \newline $\checkmark$ Can however be extended to address nonlinear interactions important when examining the effects of harmonic components \\ \hline
       $\times$ ~Requires lengthy integration times (hundreds/thousands of calculations) to account for Doppler broadening & $\checkmark$ Allows up to approximately 100 times faster computations than full numerical solution \\ \hline
    \end{tabular}
    \caption{Summary of primary tradeoffs between the standard numerical approach (e.g., ~\cite{Meyer:24}) and the LTI approach presented in this paper.}
    \label{table1}
\end{table*}
\end{center}

\section{Density Matrix Formalism}
A simple model of a realistic Rydberg receiver can be modeled by a five-energy-level structure as shown in Fig.~\ref{f:RydbergSetup}A. The atom has 4 ``primary'' states which participate in the sensor's operating process, and whose dynamics can be described by the time-dependent density matrix ${\boldsymbol{\rho}}(t)$. Representing a mixed state, the density matrix's diagonal elements define the populations of each of the energy states, while off-diagonal elements correspond to the coherences between states. We emphasize that the atomic vapor by itself is not a closed system; along with field interactions, there exist dissipative interactions with the environment that must be accounted for as well. In the model we thus introduce a fifth ``dummy state'' as a heuristic modeling tool~\cite{Sedlacekthesis}, through which populations of excited states collapse to lower-lying states via various mechanisms (\textit{e.g.}, collisional effects and other decoherences that are difficult to accurately model).

In the interaction picture, the time-evolution of the density matrix is determined by a Hamiltonian containing atom-field interactions. The coupling strengths between each pair of energy levels are written by scaling the amplitude of each electric field $\boldsymbol{E}_i$ by the dipole matrix element $\mu_i$ for each respective atomic transition; \textit{i.e.}, $\hbar\Omega_i=\vec{\mu_i}\cdot\vec{\boldsymbol{E}_i}$, where $\Omega_i$ are the associated Rabi frequencies. The dummy state is not coupled electromagnetically to any of the other states; it only provides a separate means for the atomic population to decay back to the ground state. As a result the steady-state Hamiltonian is presented as a matrix operator:
\begin{equation}
	\boldsymbol{H}=\frac{\hbar}{2}
	\begin{pmatrix}
		0 & \Omega_p & 0 & 0 & 0 \\
		\Omega_p & -2\Delta_p & \Omega_c & 0 & 0 \\
		0 & \Omega_c & -2(\Delta_p+\Delta_c) & \Omega_{RF} & 0 \\
		0 & 0 & \Omega_{RF} & -2(\Delta_p+\Delta_c+\Delta_{RF}) & 0 \\
		0 & 0 & 0 & 0 & 0
	\end{pmatrix}.
	\label{eq:hamiltonian}
\end{equation}

Here we have invoked the rotating-wave approximation (RWA) which transforms the system into a reference frame that is co-rotating with the applied fields. A consequence of removing the time-dependence is that the frequency differences between the oscillating frequency of a field and the closest atomic resonance are treated as a frequency detuning of $\Delta_i$. Note that the RWA is valid when the energy level spacings, decay rates, and driving strengths are small compared to the oscillatory frequencies of the incident fields, as is the case in a typical Rydberg atom electric field sensor. The RWA treats the atomic response to the electromagnetic fields as quasi-static and greatly simplifies the computation.

In addition to the application of external fields, environmental interactions also contribute to the atomic system evolution. These are captured in the Lindblad operator ${\boldsymbol{L}}$, where each decay or decoherence mechanism has its own term. We present the conceptual ideas here and provide additional details on the algebraic structure of each term in the Supplemental Information. In the model we consider the natural decay of each excited state ($\Gamma_{21}$, $\Gamma_{32}$) as well as decoherences between the Rydberg states leading to populating the dummy state, $\gamma_{dsg}$, through what we expect is mostly through collisional processes. In addition, we also consider dephasing associated with atoms transiting in and out of the laser fields using a parameter $\gamma_t$ that quantifies the transit dephasing rate. Finally, to close the cycle and conserve population, the ground state is repopulated at the dummy state relaxation rate $\gamma_{dsr}$. The Lindblad operator matrix is therefore defined as
\begin{equation}
	\begin{split}
		\boldsymbol{L}_{total}=\Gamma_{21}\boldsymbol{L}_{21}+\Gamma_{32}\boldsymbol{L}_{32}+\gamma_{dsg}(\boldsymbol{L}_{35}+\boldsymbol{L}_{45})\\
		+\gamma_{t}\boldsymbol{L}_{tr}+\gamma_{dsr}\boldsymbol{L}_{51}.
	\end{split}
	\label{eq:lindblad1}
\end{equation}

System evolution due to all the processes highlighted above is governed by the Liouville-Von Neumann equation~\cite{Sedlacekthesis}, given by
\begin{equation}\label{Eq:Master}
	\frac{d{\boldsymbol{\rho}}(t)}{dt} = -\frac{i}{\hbar} [{\boldsymbol{H}},{\boldsymbol{\rho}}(t)] + {\boldsymbol{L}}.
\end{equation}
Containing a set of coupled first-order ODEs, this equation is used to keep track of the time-dependent density matrix elements. However, for conducting experimental measurements using the Rydberg atom electric field sensor (where typically the measurement is conducted via photodetection of the probe field transmitted through the atoms), the quantity of interest is $\rho_{12}(t)$, whose real part determines the accumulated phase shift and whose imaginary part determines the effective susceptibility $\chi$ experienced by the probe field as it passes through the atomic vapor. These bulk properties therefore determines how the probe laser propagates through the cell. An additional RF field passing through the Rydberg atom medium leads to variations in the attenuation of probe laser power that are measured by the photodetector. Assuming the control beam power does not deplete noticeably through the cell, this implies that changes in the detector output voltage, $\delta V_{out}(t) \propto \chi(t)$ are directly related to changes in the density matrix because $\chi(t) \propto \text{Im}(\rho_{12}(t))$.

\section{Linearizing the Equations of Motion}
Based on Eq.~\ref{Eq:Master}, we recognize that all terms in the coupled equations contain a density matrix element product with a certain rate/frequency. We can therefore factor out the density matrix elements into vector form and rewrite the system of equations as a matrix operator ${\boldsymbol{A}}$ acting on a vector in the form of a flattened density matrix  $\vec{\boldsymbol{\rho}}(t)$ according to the following:
\begin{equation}
	\dot{\vec{\boldsymbol{\rho}}}(t) = {\boldsymbol{A}}\cdot\vec{\boldsymbol{\rho}}(t),
	\label{eqn:ODE}
\end{equation}
where $\vec{\boldsymbol{\rho}}(t)$ is a $25 \times 1$ vector with components $\rho_{1,1}(t),  \rho_{1,2}(t),\ldots,\rho_{1,5}(t),\rho_{2,1}(t),\ldots, \rho_{4,5}(t),\rho_{5,5}(t)$, $\dot{\vec{\boldsymbol{\rho}}}(t)$ represents its time derivative $\partial\vec{\boldsymbol{\rho}}(t)/\partial t$, and ${\boldsymbol{A}}$ is a $25 \times 25$ constant coefficient matrix independent of $\rho_{i,j}(t)$.

The steady-state solution is obtained by setting $\dot{\vec{\boldsymbol{\rho}}}(t)=0$. This solution is a vector $\vec{\boldsymbol{\rho}}(t\rightarrow\infty)\equiv\vec{\boldsymbol{\rho}}_{SS}$ that denotes the equilibrium operating point for the sensor, determined entirely by the electric field strengths, frequency detunings, and decoherence parameters included in ${\boldsymbol{A}}$.

We now consider the effect of an input RF signal field that is detuned from resonance by an intermediate frequency, $\omega_{\text{IF}}=\omega_{sig}-\omega_{LO}$. After applying the RWA, the impact of the RF signal field on the total RF field can be treated as a perturbation on the LO, written in the form of a Taylor expansion as
\begin{equation}
	\Omega_{RF}=\Omega_{RF}^{(0)} + \epsilon \Omega_{RF}^{(1)} + \epsilon^2 \Omega_{RF}^{(2)} + {\cal O}(\epsilon^3),
	\label{taylorOm}
\end{equation}
where $\Omega_{RF}^{(i)}(t)=\mu_{RF}E_{RF}^{(i)}(t)/\hbar$, $\epsilon$ is used to track the order of the Taylor expansion, and the lowest-order electric field terms can be written as
\begin{eqnarray}
  E_{RF}^{(0)} & = & E_{LO}, \\
  E_{RF}^{(1)}(t) & = & E_{LO}\text{cos}\left[\omega_{IF}(t)+\phi(t)\right],~\text{and} \\
  E_{RF}^{(2)}(t) & = & \left(E_{LO}/2\right)\text{cos}^2\left[\omega_{IF}(t)+\phi(t)\right].
\end{eqnarray}
Additionally, the induced response in the atoms can also be written as a perturbative expansion according to
\begin{equation}
	\vec{\boldsymbol{\rho}}(t) = \vec{\boldsymbol{\rho}}^{(0)} + \epsilon \vec{\boldsymbol{\rho}}^{(1)}(t) + \epsilon^2 \vec{\boldsymbol{\rho}}^{(2)}(t) + {\cal O}(\epsilon^3).
	\label{taylorRho}
\end{equation}
The zeroth order term $\vec{\boldsymbol{\rho}}^{(0)}=\vec{\boldsymbol{\rho}}_{SS}$ is the steady-state response at equilibrium. Substituting Eqs.~\ref{taylorOm} and \ref{taylorRho} into the Liouville-Von Neumann equation Eq.~\ref{Eq:Master}, we find

\begin{equation}
	\begin{aligned}
		\begin{split}
			\frac{d}{dt}(\vec{\boldsymbol{\rho}}_{SS}+ \epsilon \vec{\boldsymbol{\rho}}^{(1)}(t) +{\cal O}(\epsilon^2))=\\
			({\boldsymbol{A}}+\boldsymbol{B}(\epsilon \Omega_{RF}^{(1)}(t) +{\cal O}(\epsilon^2)))
			\cdot(\vec{\boldsymbol{\rho}}_{SS}+ \epsilon \vec{\boldsymbol{\rho}}^{(1)}(t) +{\cal O}(\epsilon^2)).
		\end{split}
	\end{aligned}
	\label{eqn:pertmaster}
\end{equation}
Because $\dot{\vec{\boldsymbol{\rho}}}_{SS}(t)=0$,
\begin{equation}
    \begin{aligned}
	\begin{split}
	    \epsilon \dot{\vec{\boldsymbol{\rho}}}^{(1)}(t) +{\cal O}(\epsilon^2)=
			{\boldsymbol{A}}\vec{\boldsymbol{\rho}}_{SS}+{\boldsymbol{A}}\epsilon \vec{\boldsymbol{\rho}}^{(1)}(t)+\\\epsilon \Omega_{RF}^{(1)}(t) {\boldsymbol{B}}\cdot\vec{\boldsymbol{\rho}}_{SS} + \epsilon^2\Omega_{RF}^{(1)}(t){\boldsymbol{B}}\cdot\vec{\boldsymbol{\rho}}^{(1)}(t)
			 +\ldots,
	\end{split}
	\end{aligned}
	\label{eqn:pertmaster2}
\end{equation}
where $\boldsymbol{B}$ is a constant-valued $25 \times 25$ sparse matrix whose non-zero elements are only those that correspond to the Rydberg-Rydberg transition. The Rabi frequency of the LO field, $\Omega_{\text{LO}}$, is  contained within ${\boldsymbol{A}}$. 

The premise of our LTI approach relies on studying the first-order (linear) response to an incident input RF signal field represented by $\Omega_{RF}^{(1)}(t)$. Collecting terms of first-order in $\epsilon$ and neglecting higher-order terms, the system of equations are separately written in matrix form as
\begin{equation}
	\begin{array}{c}
		\vec{0}=\boldsymbol{A} \cdot \vec{\boldsymbol{\rho}}_{SS}
	\end{array}
\label{eqn:SS}
\end{equation}
and
\begin{equation}
	\begin{array}{c}
		\dot{\vec{\boldsymbol{\rho}}}^{(1)}(t)= \boldsymbol{A} \cdot \vec{\boldsymbol{\rho}}^{(1)}(t) + \Omega_{RF}^{(1)}(t)\boldsymbol{B} \cdot \vec{\boldsymbol{\rho}}_{SS}.
	\end{array}
\label{eqn:FOpert}
\end{equation}
Equation~\ref{eqn:SS} is the familiar steady-state condition for the equilibrium sensing state of the receiver. Equation~\ref{eqn:FOpert} provides a way to relate changes in the atomic response $\vec{\boldsymbol{\rho}}^{(1)}(t)$ to an input RF signal field $\Omega_{RF}^{(1)}(t)$, which can be an arbitrary time-dependent incident field that we want to measure.

\section{Deriving the Transfer Matrix}
\begin{figure}[h!]
	{
		\centering
		\includegraphics[width=1.0\linewidth]{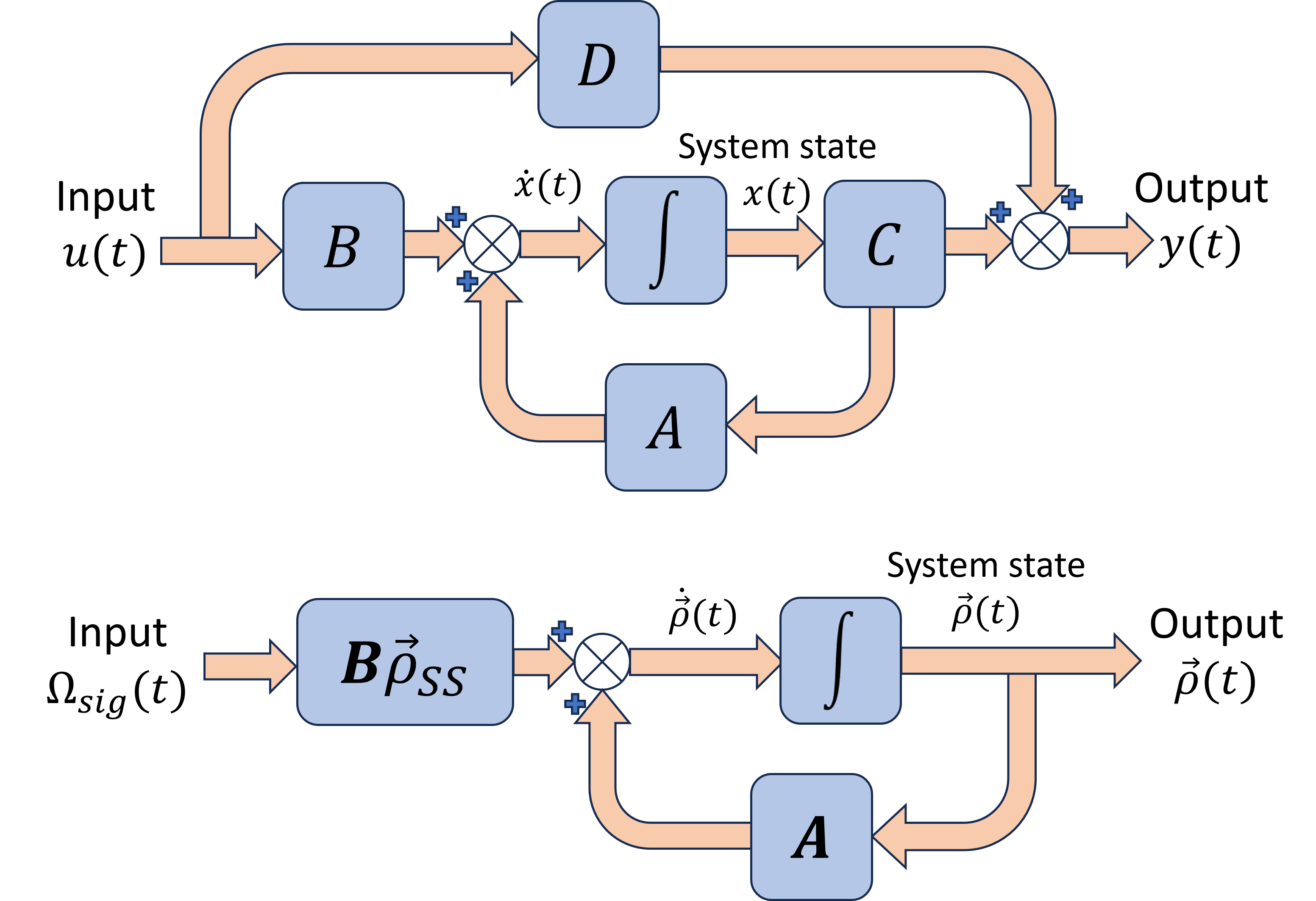}
		\caption[State-space representation of the Rydberg sensor]{(Top) A block diagram for a linear system in the state-space representation. The equivalent mathematical form is found in Eq.~\ref{eqn:statespace}. (Bottom) The entire Rydberg sensor system when linearized and put into the state-space representation. Here, the matrices $\boldsymbol{C}$ and $\boldsymbol{D}$ and their respective feedback/feedforward operations disappear, as those terms are no longer present in Eq.~\ref{eqn:pertmaster}.
		\label{f:StSpblock}}
	}
\end{figure}

Examining Eq.~\ref{eqn:FOpert} further, it can be seen that it has the same form as a state-space model of an LTI system, as depicted in Fig.~\ref{f:StSpblock}, with a state vector $\boldsymbol{x}(t)$, input $\boldsymbol{u}(t)$, and output $\boldsymbol{y}(t)$, characterized by the following set of time-dependent responses:
\begin{equation}
	\begin{array}{c}
            \dot{\boldsymbol{x}}(t)=\boldsymbol{A}(t)\boldsymbol{x}(t)+\boldsymbol{B}(t)\boldsymbol{u}(t) \\
            \boldsymbol{y}(t)=\boldsymbol{C}(t)\boldsymbol{x}(t)+\boldsymbol{D}(t)\boldsymbol{u}(t)
	\end{array}
	\label{eqn:statespace}
\end{equation}
Here, the matrices $\boldsymbol{A, B, C, D}$ manipulate and propagate the input and state vectors throughout the system and are therefore determined by the internal processes and dynamics. The input vector $\boldsymbol{u(t)}$ transforms through matrix $\boldsymbol{B}$ and is added to a transformed state vector, which determines its instantaneous rate of change. Upon integrating, the state vector is obtained, transformed by matrix $\boldsymbol{C}$ and added to any additional disturbances represented by matrix $\boldsymbol{D}$, to finally give the output or observable $\boldsymbol{y(t)}$.

LTI systems such as coupled mechanical oscillators, complex networks of circuits, and nuclear or chemical reactor controllers are typically designed to include feedback processes that act on the state of the system at a given time and keep output variables, such as power, stable \cite{hallauerLTI_2016, delchamps_1988, Vajpayee_2020, luyben_2007}. In comparison, the equations describing the Rydberg sensor system inherently and by construction contain feedback terms so that the observables, such as atomic state populations, are physically meaningful. Mapping Eq.~\ref{eqn:statespace} to Eq.~\ref{eqn:FOpert}, $\boldsymbol{C}$ and $\boldsymbol{D}$ become the identity matrix and the null matrix, respectively. This can be visualized as another diagram in the state-space representation with matrices replaced by those in Eq.~\ref{eqn:pertmaster}. Viewing the problem in a systems context is especially useful if direct comparisons are to be made between these Rydberg atom sensors and antennas or other devices used for RF field detection. Another advantage is the ability to apply commonly used analysis techniques for specifications such as sensor saturation, nonlinear behavior, and other output stability criteria.

In the remaining analyses, we will focus on extracting the frequency response of the Rydberg sensor in the linear regime, where the input RF signal field is much weaker than the LO field and Eq.~\ref{eqn:FOpert} is valid. To assess the frequency response, it is useful to transform our framework into the Fourier domain. Equation~\ref{eqn:FOpert} in the frequency domain is written as
\begin{equation}
	\begin{array}{c}
		i\omega \tilde{\vec{\boldsymbol{\rho}}}^{(1)}(\omega)= \boldsymbol{A} \cdot \tilde{\vec{\boldsymbol{\rho}}}^{(1)}(\omega) + \boldsymbol{B} \cdot \vec{\boldsymbol{\rho}}_{SS} \tilde{\Omega}_{RF}^{(1)}(\omega),
	\end{array}
	\label{eqn:FOfourier}
\end{equation}
where $\tilde{\vec{\boldsymbol{\rho}}}^{(1)}(\omega)$ and $\tilde{\Omega}_{RF}^{(1)}(\omega)$ are the Fourier transforms of $\vec{\boldsymbol{\rho}}^{(1)}(t)$ and $\Omega_{RF}^{(1)}(t)$, respectively. By factoring and isolating the output response term, it can be shown that
\begin{equation}
	\begin{array}{c}
		(i\omega\boldsymbol{I}-\boldsymbol{A})\cdot \tilde{\vec{\boldsymbol{\rho}}}^{(1)}(\omega)=  \boldsymbol{B}\cdot \vec{\boldsymbol{\rho}}_{SS} \tilde{\Omega}_{RF}^{(1)}(\omega), \text{~and} \\
		 \\
		\tilde{\vec{\boldsymbol{\rho}}}^{(1)}(\omega)= (i\omega\boldsymbol{I}-\boldsymbol{A})^{-1} \boldsymbol{B}\cdot \vec{\boldsymbol{\rho}}_{SS} \tilde{\Omega}_{RF}^{(1)}(\omega),
	\end{array}
\label{eqn:FOfourier2}
\end{equation}
where $\boldsymbol{I}$ is the $25 \times 25$ identity matrix. Equation~\ref{eqn:FOfourier2} provides a convenient way to describe the linear output response as a frequency-dependent function of the input field. This notation is very similar to transfer functions derived in the state-space formulation and control theory in general, where the output(s) $Y$ of a system are related to the input(s) $U$ by
\begin{equation}
	Y(\omega) = G(\omega)U(\omega),
	\label{eqn:TFinStSp}
\end{equation}
and $G(\omega)$ is the linear frequency transfer function of that system. We therefore write the Rydberg atom sensor response as
\begin{equation}
	\begin{array}{c}
		\tilde{\vec{\boldsymbol{\rho}}}^{(1)}(\omega)=\boldsymbol{G}(\omega) \tilde{\Omega}_{RF}^{(1)}(\omega)
	\end{array}
	\label{eqn:TF1}
\end{equation}
where $\boldsymbol{G}(\omega)\equiv(i\omega\boldsymbol{I}-\boldsymbol{A})^{-1}\boldsymbol{B} \cdot \vec{\boldsymbol{\rho}_{SS}}$ is another $25 \times 25$ matrix. Note that the vector $\tilde{\vec{\boldsymbol{\rho}}}^{(1)}(\omega)$ represents the frequency response of all elements of the density matrix.  In particular, the Fourier transform of $Im(\rho_{1,2}^{(1)}(t))$ is proportional to the probe field transmission through the vapor cell. Taking the matrix elements associated with $\boldsymbol{\rho_{1,2}}(\omega)$ and $\boldsymbol{\rho_{2,1}}(\omega)$ in $\boldsymbol{G}$ and using the identity $Im(\boldsymbol{Z}) = (\boldsymbol{Z} - \boldsymbol{Z^*})/2i$, we can write a transfer function for probe laser transmission in the intermediate frequency $\omega_{IF}$ in the following compact manner:
\begin{equation}
	\begin{array}{c}
		 \tilde{\rho}^{(1)}(\omega_{IF}) = G(\omega_{IF}) \tilde{\Omega}_{sig}(\omega_{IF}),
	\end{array}
	\label{eqn:TF3}
\end{equation}
where 
\begin{equation}
    \begin{array}{c}
        G(\omega_{IF}) \equiv [\boldsymbol{G}_{1,2}(\omega)-\boldsymbol{G}_{2,1}(\omega)]/2i 
    \end{array}
\end{equation}
is analogous to its counterpart in Eq.~\ref{eqn:TFinStSp}.

\subsection{Incorporating Doppler Averaging}
The analysis above is performed for one set of probe and control field detunings, $\Delta_p$ and $\Delta_c$, which are contained in $\boldsymbol{A}$. To include finite temperature effects where the atoms are moving around in the vapor cell, Doppler-induced laser detunings must be taken into account. We add these as Doppler shifts from velocities along the optical field propagation that are sampled from a thermal Maxwell distribution according to
\begin{equation}
	f(u)=\frac{1}{\sqrt{\pi}}\int_{-\infty}^{\infty}e^{-u^2}du,
	\label{eqn:maxwell}
\end{equation}
where $u\equiv{v/\sigma_v}$, $\sigma_v=\sqrt{2 k_B T/m}$, $v$ is the atom speed, $k_B$ is Boltzmann's constant, $T$ is the average atomic temperature, $m$ is the atomic mass. The effective detunings of the probe and control lasers for any velocity class $u$, set by the root-mean-square velocity for the thermal distribution, are given by the following:
\begin{equation}
	\begin{array}{c}
		\Delta_{p}^{'}(u)=\Delta_{p,0}+\frac{2\pi\sigma_v}{\lambda_{p}}u \\
		\\
		\Delta_{c}^{'}(u)=\Delta_{c,0}-\frac{2\pi\sigma_v}{\lambda_{c}}u.
	\end{array}
	\label{eqn:effdets}
\end{equation}
Here, $\Delta_{p,0}$ and $\Delta_{c,0}$ are the overall detunings of the probe and control fields from the atomic resonances and $\lambda_{p}$ and $\lambda_{c}$ are their wavelengths, respectively. By making the detuning dependence in the matrix $\boldsymbol{G}$ explicit, the transfer function becomes
\begin{multline}
		G(\omega;u)=
		\frac{1}{2i}[\boldsymbol{G}_{1,2}(\Delta_{p}^{'}(u),\Delta_{c}^{'}(u))-\boldsymbol{G}_{2,1}(\Delta_{p}^{'}(u),\Delta_{c}^{'}(u)))].
	\label{eqn:TFdopp}
\end{multline}
To obtain the Doppler-averaged transfer matrix $\bar{G}(\omega)$, we need to integrate over all velocity classes according to
\begin{equation}
	\bar{G}(\omega)=\frac{1}{\sqrt{\pi}}\int_{-\infty}^{\infty}e^{-u^2}G(\omega;u)du.
	\label{eqn:TFintfinal}
\end{equation}

\begin{figure}[h!]
	{
		\centering
		\includegraphics[width=1.0\linewidth]{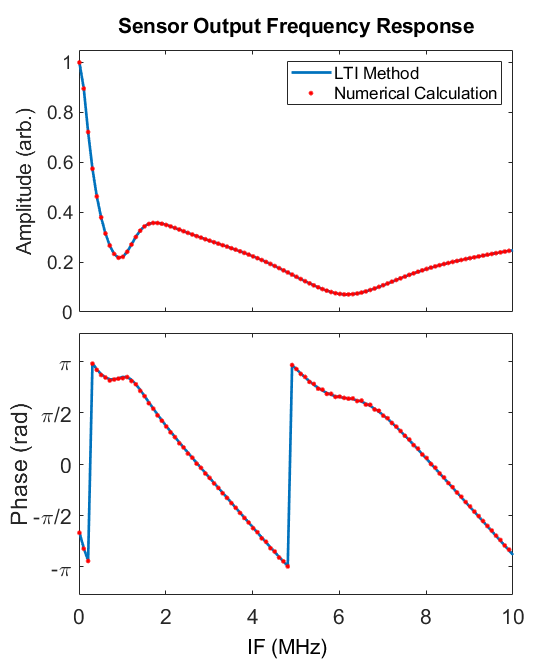}
		\caption[Example Transfer Function]{An example of a transfer function for a 4+1 level Rydberg sensor, as described in the main text. (Top) Normalized amplitude response of the probe laser transmission. The solid blue line shows the response calculated using the LTI method, obtained by taking the magnitude of the complex-valued transfer function at each intermediate frequency. The pink dots are obtained by fitting the last 5 oscillations of the full time-dependent numerical solution when a small modulated signal is applied. (Bottom) The phase response that is obtained by taking the argument of the complex-valued transfer function. The phase has been wrapped and shifted vertically to ensure that the fitted values of the phase lie between -$\pi$ and $\pi$. All simulations were conducted using the parameters summarized in Table~\ref{tableparams}. 
		\label{f:TFplot}}
	}
\end{figure}

For the vast majority of Rydberg sensor schemes, the physical observable is the probe laser power that is transmitted through the atomic vapor cell. Using Beer's Law, $P/P_0=e^{-\alpha L}$, the ratio of received power $P$ to incident power $P_0$ is determined by the absorption coefficient $\alpha$ after passing through a sensing volume of length $L$. Since $\alpha \propto\ $Im$(\rho_{12})$, we focus our attention on this density matrix element. Figure \ref{f:TFplot} shows the response of the probe transmission as a function of the modulation frequency (\textit{i.e.}, the transfer function of the observed signal output of the sensor). 

We compute the transfer function with Doppler averaging using our LTI method described above for 101 values of the intermediate frequency ranging from $0$ to $10$~MHz, shown in solid blue, assuming an amplitude-modulated signal. We also perform a full numerical solution of the transfer function by integrating the raw equations of motion in time using an open-source state-of-the-art Rydberg sensor modeling package called RydIQule~\cite{Meyer:24}. In this calculation we manually insert the steady-state field couplings and decay parameters, applying a modulation at each frequency (red dots), and then solve for the Doppler-averaged time-dependent density matrix. Fitting the time-domain element Im$(\rho_{12}(t))$ to a sinusoid, we extract the amplitude information and display the transfer function results. While the full numerical integration took about 2 hours, our LTI system approach took a little over a minute. This large speedup is particularly noticeable for the smaller intermediate frequencies where longer oscillation periods are needed for accurate fitting of the numerical sinusoidal waveform.

\begin{figure}[h!]
    \centering
    \begin{minipage}{0.45\textwidth}
    \centering
    \includegraphics[scale=0.4]{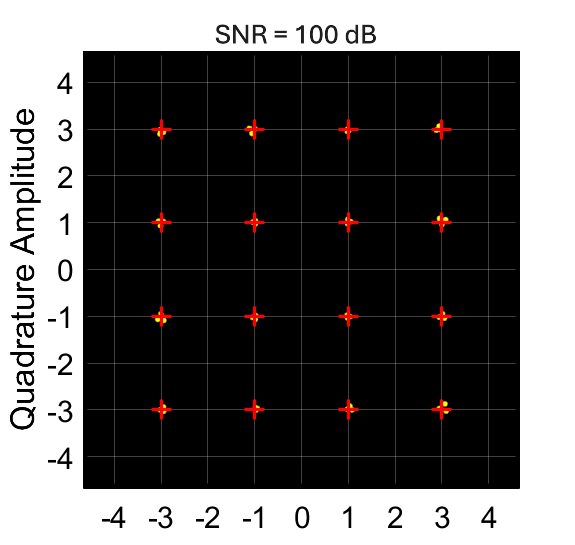}
    \end{minipage}\vfill
    \begin{minipage}{0.45\textwidth}
    \centering
    \includegraphics[scale=0.4]{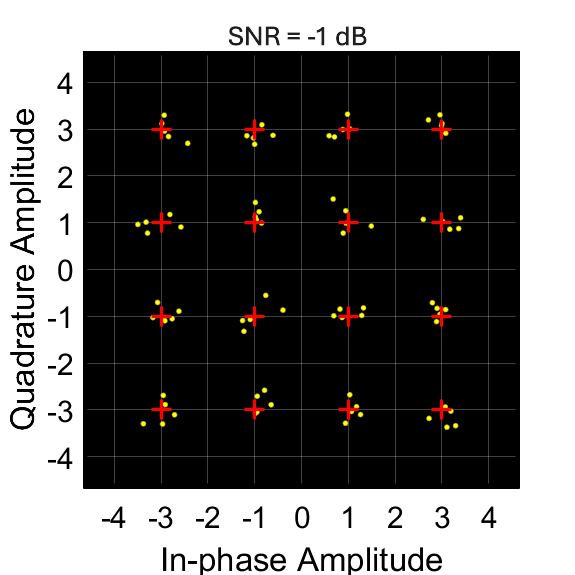}
    \end{minipage}
    \caption{Analysis of the reception of a 16QAM waveform. An example transmitted waveform containing all 16 components is received by the Rydberg sensor and its response is calculated using our LTI method. Here, each symbol contained in the waveform is a mapping of a transmitted bit of information onto the in-phase and quadrature components. The simulated sensor output is fit to a sinusoid to extract the amplitude and phase; the received waveform is reconstructed and presented in a constellation diagram. (Top) When the sensor receives a signal with a very high signal-to-noise ratio, reconstruction between the transmitted signal (red dots) and received signal (yellow dots) is robust, and contained information can be extracted with a high accuracy. (Bottom) Instead, if an incoming RF signal has a large noise component, the reconstructed signal has a larger variance. Details of the simulation can be found in Appendix B.} 
    \label{fig:QAMexample}
\end{figure}

While the transfer function provided in Fig.~\ref{f:TFplot} describes the system response for an amplitude-modulated input signal field, an analogous manipulation of the LTI framework can be used to obtain a transfer function for other modulation schemes as well. As one example, Fig.~\ref{fig:QAMexample} provides an example of a quadrature amplitude modulation scheme (16QAM), whose 16 states are best represented as a constellation diagram. Each state is defined by its own amplitude and phase and is marked by red crosses. To generate these responses, a signal containing 5 repetitions of all components in a 16QAM waveform is simulated in the time domain. To ensure that all the individual symbols have time to reach steady-state oscillatory behavior the entire waveform lasts for over $1.5~ms$, compared to the several microsecond-scale signals used to calculate the red points in Fig.~\ref{f:TFplot}. This means that a brute-force time-dependent integration of Eq.~\ref{Eq:Master} would take an unreasonably long time for a full 16QAM waveform sensor response. 

The 16QAM waveform is then ``passed" through the Rydberg sensor by convolving it with its LTI system impulse response function. This output response can then be compared to the input to quantify the sensor's reception fidelity for the waveform. To obtain the yellow dots, the output is fit to a sinusoid, its amplitude and phase are extracted, and its quadratures plotted on the constellation diagram. For a ``clean" input QAM waveform the sensor produces a output response that is extremely close to the input when reconstructed (top plot in Fig.~\ref{fig:QAMexample}).

Because the phase response of the Rydberg sensor is not linear, a short signal pulse suffers from some dispersion~\cite{PhysRevApplied.18.034030}, which can add noise to the sensor's response. For example, the constellation diagrams for phase-shift keying or QAM protocols may appear blurry due to low SNR and/or the sensor's nonlinear response. Our LTI method can be applied to rapidly evaluate the Rydberg sensor response to various waveforms and modulation frequencies to better understand these behaviors, as visualized for an SNR$~=-1$ dB 16QAM signal in Fig.~\ref{fig:QAMexample} (bottom figure). Details of the waveform analysis are highlighted in Apprendix~\ref{appB}. Here the results show up on the constellation diagram as deviations from a 16QAM signal and whose fidelity can be quantified by an error-vector magnitude (EVM). The EVM represents the quality of the received modulated waveform compared to an ideal decoding of the waveform. The vector from a given red point to a yellow dot is the error in the magnitude and phase of a symbol in the 16QAM waveform, and the relative distance between the two is defined by the EVM. In practice, waveform reception is not only a product of noise but also distortions such as frequency noise, AM-AM and AM-PM distortions among other sources.

\section{Conclusion}
We have presented a technique to quickly obtain the frequency response for a density matrix that describes the state of the Rydberg atom electric field sensor. Using the Liouville-von-Neumann state equations derived from first principles, we linearize the Rydberg sensor state variables and frame the equations of motion as a systems engineering problem. Armed with the Rydberg sensor-LTI system mapping method, analysis of the sensor's linear regime naturally follows, and time-domain calculations for responses to arbitrary waveforms can be easily obtained via the impulse response function. This approach can be used to conduct predictions of key sensor performance metrics, such as sensitivity and dynamic range, for a given experimental configuration. As part of future work, moving beyond the linear regime, we plan to extend our method to study nonlinear behavior where metrics such as dynamic range and intercept points are of much interest in broader engineering problems.


\section*{Acknowledgements}
The authors acknowledge funding from the MITRE Independent Research and Development Program and the Defense Advanced Research Projects Agency (DARPA) under Contract No.~W56KGU-18-D-0004/W56KGU-21-F-0008. The views, opinions, and/or findings expressed are those of the author(s) and should not be interpreted as representing the official views or policies of the Department of Defense or the U.S. Government. Distribution Statement "A" (Approved for Public Release, Distribution Unlimited) \copyright2025 The MITRE Corporation. ALL RIGHTS RESERVED. Approved for Public Release; Distribution Unlimited. Public Release Case Number 25-0332.

\section*{References}
\bibliography{bibliography}

\newpage
\begin{widetext}
\label{appA}
\section{Appendix A - Full Expressions for Terms in the Equations of Motion}
For the system described in the main text, the following set of equations can be used to define the matrix $\boldsymbol{A}$ (introduced in Eq.~\ref{eqn:ODE}):

\begin{equation}
\boldsymbol{A}=
 \begin{pmatrix}
    \mathbb{A}_{11} & \mathbb{A}_{12} & \mathbb{A}_{13} & \mathbb{A}_{14} & \mathbb{A}_{15} \\
    \mathbb{A}_{21} & \mathbb{A}_{22} & \mathbb{A}_{23} & \textbf{0} & \textbf{0} \\
    \textbf{0} & \mathbb{A}_{32} & \mathbb{A}_{33} & \mathbb{A}_{34} & \textbf{0} \\
    \textbf{0} & \textbf{0} & \mathbb{A}_{43} & \mathbb{A}_{44} & \textbf{0} \\
    \textbf{0} & \textbf{0} & \mathbb{A}_{53} & \mathbb{A}_{54} & \mathbb{A}_{55} \\
 \end{pmatrix},
 \end{equation}
 where $\textbf{0}$ is a $5\times5$ matrix of zeroes,
\begin{equation}
\mathbb{A}_{11}=
 \begin{pmatrix}
    0 & -\frac{i\Omega_p}{2} & 0 & 0 & 0 \\
    -\frac{i\Omega_p}{2} & -\frac{\Gamma_{21}}{2}-i\Delta_p & -\frac{i\Omega_c}{2} & 0 & 0 \\
    0 & -\frac{i\Omega_c}{2} & -\frac{\left(\tilde{\gamma}+\Gamma_{32}\right)}{2}-i\left(\Delta_p+\Delta_c\right) & -\frac{i\Omega_{\text{RF}}}{2} & 0 \\
    0 & 0 & -\frac{i\Omega_{\text{RF}}}{2} & -\frac{\tilde{\gamma}}{2}-i\left(\Delta_p+\Delta_c+\Delta_{RF}\right) & 0 \\
    0 & 0 & 0 & 0 & -\frac{\gamma_{dsr}}{2} \\
 \end{pmatrix},
 \end{equation}
 
\begin{equation}
\mathbb{A}_{12}=
 \begin{pmatrix}
    \frac{i\Omega_p}{2} & \Gamma_{21} & 0 & 0 & 0 \\
    0 & \frac{i\Omega_p}{2} & 0 & 0 & 0 \\
    0 & 0 & \frac{i\Omega_p}{2} & 0 & 0 \\
    0 & 0 & 0 & \frac{i\Omega_p}{2} & 0 \\
    0 & 0 & 0 & 0 & \frac{i\Omega_p}{2} \\
 \end{pmatrix},
 \end{equation}
 
\begin{equation}
\mathbb{A}_{13}=
 \begin{pmatrix}
    0 & 0 & \gamma_t & 0 & 0 \\
    0 & 0 & 0 & 0 & 0 \\
    0 & 0 & 0 & 0 & 0 \\
    0 & 0 & 0 & 0 & 0 \\
    0 & 0 & 0 & 0 & 0 \\
 \end{pmatrix},
 \end{equation}
 
\begin{equation}
\mathbb{A}_{14}=
 \begin{pmatrix}
    0 & 0 & 0 & \gamma_t & 0 \\
    0 & 0 & 0 & 0 & 0 \\
    0 & 0 & 0 & 0 & 0 \\
    0 & 0 & 0 & 0 & 0 \\
    0 & 0 & 0 & 0 & 0 \\
 \end{pmatrix},
 \end{equation}
 
\begin{equation}
\mathbb{A}_{15}=
 \begin{pmatrix}
    0 & 0 & 0 & 0 & \gamma_{dsr} \\
    0 & 0 & 0 & 0 & 0 \\
    0 & 0 & 0 & 0 & 0 \\
    0 & 0 & 0 & 0 & 0 \\
    0 & 0 & 0 & 0 & 0 \\
 \end{pmatrix},
 \end{equation}

 
  \begin{equation}
\mathbb{A}_{21}=
 \begin{pmatrix}
    \frac{i\Omega_p}{2} & 0 & 0 & 0 & 0 \\
    0 & \frac{i\Omega_p}{2} & 0 & 0 & 0 \\
    0 & 0 & \frac{i\Omega_p}{2} & 0 & 0 \\
    0 & 0 & 0 & \frac{i\Omega_p}{2} & 0 \\
    0 & 0 & 0 & 0 & \frac{i\Omega_p}{2} \\
 \end{pmatrix},
 \end{equation}
 
\begin{equation}
\mathbb{A}_{22}=
 \begin{pmatrix}
    -\frac{\Gamma_{21}}{2}+i\Delta_p & -\frac{i\Omega_p}{2} & 0 & 0 & 0 \\
    -\frac{i\Omega_p}{2} & -\Gamma_{21} & -\frac{i\Omega_c}{2} & 0 & 0 \\
    0 & -\frac{i\Omega_c}{2} & -\frac{\left(\tilde{\gamma}+\Gamma_{21}+\Gamma_{32}\right)}{2}-i\Delta_c & -\frac{i\Omega_{\text{RF}}}{2} & 0 \\
    0 & 0 & -\frac{i\Omega_{\text{RF}}}{2} & -\frac{\left(\tilde{\gamma}+\Gamma_{21}\right)}{2}-i\left(\Delta_c+\Delta_{RF}\right) & 0 \\
    0 & 0 & 0 & 0 & -\frac{\left(\gamma_{dsr}+\Gamma_{21}\right)}{2}+i\Delta_p \\
 \end{pmatrix},
 \end{equation}
 
 \begin{equation}
\mathbb{A}_{23}=
 \begin{pmatrix}
    \frac{i\Omega_c}{2} & 0 & 0 & 0 & 0 \\
    0 & \frac{i\Omega_c}{2} & \Gamma_{32} & 0 & 0 \\
    0 & 0 & \frac{i\Omega_c}{2} & 0 & 0 \\
    0 & 0 & 0 & \frac{i\Omega_c}{2} & 0 \\
    0 & 0 & 0 & 0 & \frac{i\Omega_c}{2} \\
 \end{pmatrix},
 \end{equation}
 
\begin{equation}
\mathbb{A}_{32}=
 \begin{pmatrix}
    \frac{i\Omega_c}{2} & 0 & 0 & 0 & 0 \\
    0 & \frac{i\Omega_c}{2} & 0 & 0 & 0 \\
    0 & 0 & \frac{i\Omega_c}{2} & 0 & 0 \\
    0 & 0 & 0 & \frac{i\Omega_c}{2} & 0 \\
    0 & 0 & 0 & 0 & \frac{i\Omega_c}{2} \\
 \end{pmatrix},
\end{equation}

\begin{equation}
\mathbb{A}_{33}=
 \begin{pmatrix}
    -\frac{\left(\tilde{\gamma}+\Gamma_{32}\right)}{2}+i\left(\Delta_p+\Delta_c\right) & -\frac{i\Omega_p}{2} & 0 & 0 & 0 \\
    -\frac{i\Omega_p}{2} & -\frac{\left(\tilde{\gamma}+\Gamma_{21}+\Gamma_{32}\right)}{2}+i\Delta_c & -\frac{i\Omega_c}{2} & 0 & 0 \\
    0 & -\frac{i\Omega_c}{2} & -\tilde{\gamma}-\Gamma_{32} & -\frac{i\Omega_{\text{RF}}}{2} & 0 \\
    0 & 0 & -\frac{i\Omega_{\text{RF}}}{2} & -\gamma_{dsg}-\gamma_t-\frac{\Gamma_{32}}{2}-i\Delta_{\text{RF}} & 0 \\
    0 & 0 & 0 & 0 & -\frac{\left(\tilde{\gamma}+\gamma_{dsr}+\Gamma_{32}\right)}{2}+i\left(\Delta_p+\Delta_c\right) \\
 \end{pmatrix},
 \end{equation}

\begin{equation}
\mathbb{A}_{34}=\mathbb{A}_{43}=
 \begin{pmatrix}
    \frac{i\Omega_{\text{RF}}}{2} & 0 & 0 & 0 & 0 \\
    0 & \frac{i\Omega_{\text{RF}}}{2} & 0 & 0 & 0 \\
    0 & 0 & \frac{i\Omega_{\text{RF}}}{2} & 0 & 0 \\
    0 & 0 & 0 & \frac{i\Omega_{\text{RF}}}{2} & 0 \\
    0 & 0 & 0 & 0 & \frac{i\Omega_{\text{RF}}}{2} \\
 \end{pmatrix},
\end{equation}

\begin{multline}
\mathbb{A}_{44}=\\
 \begin{pmatrix}
    -\frac{\tilde{\gamma}}{2}+i\left(\Delta_p+\Delta_c+\Delta_{\text{RF}}\right) & -\frac{i\Omega_p}{2} & 0 & 0 & 0 \\
    -\frac{i\Omega_p}{2} & -\frac{\left(\tilde{\gamma}+\Gamma_{21}\right)}{2}+i\left(\Delta_c+\Delta_{\text{RF}}\right) & -\frac{i\Omega_c}{2} & 0 & 0 \\
    0 & -\frac{i\Omega_c}{2} & -\tilde{\gamma}-\frac{\Gamma_{32}}{2}+i\Delta_{\text{RF}} & -\frac{i\Omega_{\text{RF}}}{2} & 0 \\
    0 & 0 & -\frac{i\Omega_{\text{RF}}}{2} & -\gamma_{dsg}-\gamma_t & 0 \\
    0 & 0 & 0 & 0 & -\frac{\left(\tilde{\gamma}+\gamma_{dsr}\right)}{2}+i\left(\Delta_p+\Delta_c+\Delta_{\text{RF}}\right) \\
 \end{pmatrix},
 \end{multline}

\begin{equation}
\mathbb{A}_{53}=
 \begin{pmatrix}
    0 & 0 & 0 & 0 & 0 \\
    0 & 0 & 0 & 0 & 0 \\
    0 & 0 & 0 & 0 & 0 \\
    0 & 0 & 0 & 0 & 0 \\
    0 & 0 & \gamma_{dsg} & 0 & 0 \\
 \end{pmatrix},
 \end{equation}

\begin{equation}
\mathbb{A}_{54}=
 \begin{pmatrix}
    0 & 0 & 0 & 0 & 0 \\
    0 & 0 & 0 & 0 & 0 \\
    0 & 0 & 0 & 0 & 0 \\
    0 & 0 & 0 & 0 & 0 \\
    0 & 0 & 0 & \gamma_{dsg} & 0 \\
 \end{pmatrix},
 \end{equation}
 and
\begin{equation}
\mathbb{A}_{55}=
 \begin{pmatrix}
    -\frac{\gamma_{dsr}}{2} & -\frac{i\Omega_p}{2} & 0 & 0 & 0 \\
    -\frac{i\Omega_p}{2} & -\frac{\left(\gamma_{dsr}+\Gamma_{21}\right)}{2}-i\Delta_p & -\frac{i\Omega_c}{2} & 0 & 0 \\
    0 & -\frac{i\Omega_c}{2} & -\frac{\left(\tilde{\gamma}+\gamma_{dsr}-\Gamma_{32}\right)}{2}-i\left(\Delta_p+\Delta_c\right) & -\frac{i\Omega_{\text{RF}}}{2} & 0 \\
    0 & 0 & -\frac{i\Omega_{\text{RF}}}{2} & -\frac{\left(\tilde{\gamma}+\gamma_{dsr}\right)}{2}-i\left(\Delta_p+\Delta_c+\Delta_{\text{RF}}\right) & 0 \\
    0 & 0 & 0 & 0 & -\gamma_{dsr} \\
 \end{pmatrix},
 \end{equation}
with $\tilde{\gamma}=\gamma_{dsg}+\gamma_t$.

Additionally, the following set of equations can be used to define the matrix $\boldsymbol{B}$ (introduced in Eq.~\ref{eqn:pertmaster2}):
\begin{equation}
\boldsymbol{B}=
 \begin{pmatrix}
    \mathbb{B}_{11} & \textbf{0} & \textbf{0} & \textbf{0} & \textbf{0} \\
    \textbf{0} & \mathbb{B}_{22} & \textbf{0} & \textbf{0} & \textbf{0} \\
    \textbf{0} & \textbf{0} & \mathbb{B}_{33} & \mathbb{B}_{34} & \textbf{0} \\
    \textbf{0} & \textbf{0} & \mathbb{B}_{43} & \mathbb{B}_{44} & \textbf{0} \\
    \textbf{0} & \textbf{0} & \textbf{0} & \textbf{0} & \mathbb{A}_{55} \\
 \end{pmatrix},
 \end{equation}
where
\begin{equation}
\mathbb{B}_{11}=\mathbb{B}_{22}=\mathbb{B}_{33}=\mathbb{B}_{44}=\mathbb{B}_{55}=
 \begin{pmatrix}
    0 & 0 & 0 & 0 & 0 \\
    0 & 0 & 0 & 0 & 0 \\
    0 & 0 & 0 & -\frac{i}{2} & 0 \\
    0 & 0 & -\frac{i}{2} & 0 & 0 \\
    0 & 0 & 0 & 0 & 0 \\
 \end{pmatrix},
\end{equation}
and
\begin{equation}
\mathbb{B}_{34}=\mathbb{B}_{43}=
 \begin{pmatrix}
    \frac{i}{2} & 0 & 0 & 0 & 0 \\
    0 & \frac{i}{2} & 0 & 0 & 0 \\
    0 & 0 & \frac{i}{2} & 0 & 0 \\
    0 & 0 & 0 & \frac{i}{2} & 0 \\
    0 & 0 & 0 & 0 & \frac{i}{2} \\
 \end{pmatrix}.
\end{equation}
\vspace{10mm}
\end{widetext}

\section{Appendix B - Simulation parameters}
\label{appB}
\subsection{Transfer function model}
For both the simulations shown in Fig.~\ref{f:TFplot}, the 4-level model with a dummy state includes couplings among 4(+1) states by 3 electromagnetic fields. States $\left|1\right>$ and $\left|2\right>$ are coupled by the probe laser field with a Rabi frequency, $\Omega_p$, which is proportional to the electric field amplitude. Similarly, the control laser and local oscillator fields, which couple states $\left|2\right>\leftrightarrow\left|3\right>$ and $\left|3\right>\leftrightarrow\left|4\right>$, respectively, have Rabi frequencies of $\Omega_c$ and $\Omega_{\text{LO}}$. Each of these fields has its own detuning from the transition resonant frequency (i.e. $\Delta_p,~\Delta_c,~\Delta_{\text{LO}}$). Furthermore, decoherence mechanisms play a role in the system evolution through the Lindblad operator, as in Eq.~\ref{eq:lindblad1}. All values chosen for these variables are summarized in Table~\ref{tableparams}.

\begin{table}[b]
\caption{Parameters used in the atomic system model.}
    \begin{tabular}{|c|c|}
    \hline
      \textbf{Parameter}   & 
     \textbf{Value} \\ \hline
       $\Omega_p$  & $2\pi*(7.5~\text{MHz})$ \\ \hline
       $\Omega_c$ & $2\pi*(7.5~\text{MHz})$ \\ \hline
       $\Omega_{\text{LO}}$ & $2\pi*(15~\text{MHz})$ \\ \hline
       $\Delta_{p}$ & $2\pi*(0~\text{MHz})$ \\ \hline
       $\Delta_{c}$ & $2\pi*(0~\text{MHz})$ \\ \hline
       $\Delta_{\text{LO}}$ & $2\pi*(0~\text{MHz})$ \\ \hline
       $\Gamma_{21}$ & $2\pi*(6~\text{MHz})$ \\ \hline
       $\Gamma_{32}$ & $2\pi*(1~\text{kHz})$ \\ \hline
       $\Gamma_{t}$ & $2\pi*(0.1~\text{MHz})$ \\ \hline
       $\gamma_{dsg}$ & $2\pi*(0.5~\text{MHz})$ \\ \hline
       $\Gamma_{dsr}$ & $2\pi*(0.1~\text{MHz})$ \\ \hline
    \end{tabular}
    \label{tableparams}
\end{table}

\begin{figure}
    \centering
    \begin{minipage}{0.45\textwidth}
    \centering
    \includegraphics[scale=0.5]{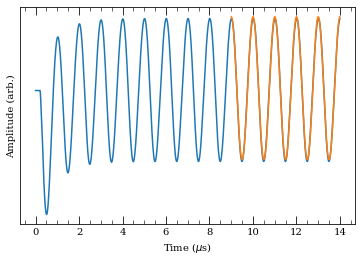}
    \end{minipage}\vfill
    \begin{minipage}{0.45\textwidth}
    \centering
    \includegraphics[scale=0.5]{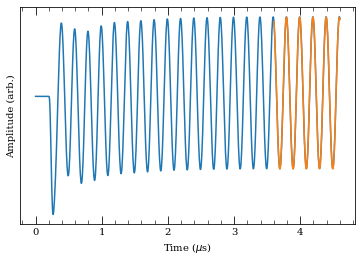}
    \end{minipage}
    \caption{Examples of time-domain responses to a weak input signal field. An RF field with one-hundredth the strength of the local oscillator is modulated onto it. The resulting equations of motion are integrated numerically using RydIQule for two select intermediate frequencies (1~MHz for the top plot and 5~MHz for the bottom plot). Blue curves are the full solutions starting at $t~=0$, while the orange curves are fit functions whose amplitudes are plotted as the red points in Fig.~\ref{f:TFplot}. Note: The curves shown here are a segment of the actual numerical calculation for better visualization of the transient features that are a signature of atomic level dephasing before reaching a steady-state driven oscillator-like behavior.}
    \label{fig:timedep}
\end{figure}

The analytical LTI approach outlined in the main text is used to obtain the blue curve in Fig.~\ref{f:TFplot}. We normalize the amplitude response so that the DC response is set to unity. For the numerical calculations we use RydIQule\textemdash a state-of-the-art Rydberg atom sensor modeling package~\cite{Meyer:24}. We set up an atomic energy level system with the above parameters and first calculate the imaginary component of the $\rho_{12}$ steady-state density matrix. Next, we add a time-dependent RF signal whose electric field amplitude is 100 times weaker than that of the local oscillator\textemdash thus imposing a weak amplitude modulation on the RF carrier\textemdash and compute the same density matrix. Two examples of the resulting time-dependent transmissions are shown in Fig.~\ref{fig:timedep} (zoomed in to show the cyclic and transient features). Once the oscillation amplitudes reach equilibrium, the last 5 cycles (shown as orange) are fit to a sinusoidal function, and the amplitude at each intermediate frequency is extracted.  

\subsection{Analysis of a 16QAM waveform}
To demonstrate the effectiveness of the LTI analysis approach, we use an example time-domain waveform and propagate it through the transfer function to yield the sensor's response. We choose a 16QAM waveform that consists of 5 repetitions of 16 ``symbols''\textemdash each with its own amplitude and phase relative to the local oscillator. The waveform is better illustrated on a constellation diagram, such as the one shown in Fig.~\ref{fig:QAMexample}, where the red dots determine the symbol's in-phase and quadrature components.

\begin{figure}[h!]
    \centering
    \includegraphics[scale=0.4]{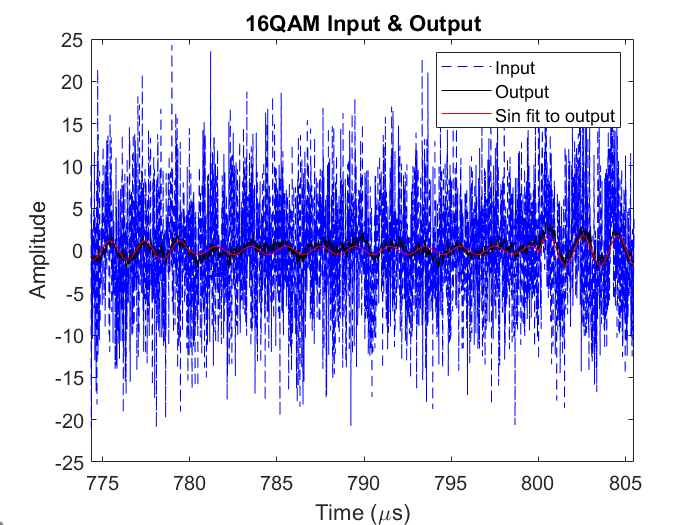}
    \caption{A segment of a 16QAM in the time-domain with a large noise amplitude. The entire waveform (dashed blue line) is fed into the sensor's LTI model, where the output signal (solid black line) is produced via the system's transfer function. Next, each symbol's amplitude and phase are obtained with a fit function (solid red line). Only a short portion of the entire waveform is shown to highlight the features of the input, output and noise features.}
    \label{fig:QAMtimedep_full}
\end{figure}

We create a time-dependent function with each symbol lasting long enough that steady-state behavior is established, as shown in Fig.~\ref{fig:QAMtimedep_full}. The total duration of the signal is over $1.5~\text{ms}$, which is extremely long for a full numerical calculation of the response to be performed for practical situations.

To analyze the 16QAM waveform, we determine the fidelity of the sensor's performance by comparing the output response to the input signal waveform. Again, by applying a sinusoidal fit, the amplitude and phase are extracted and overlayed onto the top constellation diagram in Fig.~\ref{fig:QAMexample}. Note that this example is idealized and not realistic in many practical environments. As an extension of the LTI model's versatility, we create a second ``noisy'' 16QAM waveform that has a signal-to-noise ratio of $-1$~dB (see Fig.~\ref{fig:QAMtimedep_full}). When reconstructed the noise clearly shows up on the constellation diagram (bottom plot in Fig.~\ref{fig:QAMexample}) as deviations of the reconstructed yellow points from the input red points.

\end{document}